# Single Top Quarks at the Tevatron

A.P. Heinson
*University of California, Riverside, CA 92521, USA*

For the CDF and DØ Collaborations

After many years searching for electroweak production of top quarks [1], the Tevatron collider experiments have now moved from obtaining first evidence for single top quark production to an impressive array of measurements that test the standard model in several directions. This paper describes measurements of the single top quark cross sections, limits set on the CKM matrix element $|V_{tb}|$, searches for production of single top quarks produced via flavor-changing neutral currents and from heavy $W'$ and $H^+$ boson resonances, and studies of anomalous $Wtb$ couplings. It concludes with projections for future expected significance as the analyzed datasets grow.

## 1. SINGLE TOP QUARK PRODUCTION

Until the Large Hadron Collider at CERN starts operating, the Tevatron collider at Fermi National Accelerator Laboratory is the only place where top quarks can be produced and studied. The Tevatron accelerates and collides protons on antiprotons at a center-of-mass energy of 1.96 TeV every 396 ns, which creates about 50 top-quark – top-antiquark pairs on average in a day's running. The single top quark cross section is predicted in the standard model (SM) to be about half that for $t\bar{t}$ pair production, with two main modes of generation: the s-channel process $q\bar{q}' \to t\bar{b}$, and the t-channel process $q'g \to tq\bar{b}$, plus charge-conjugate processes. For simplicity, we refer to the s-channel process as "$tb$," including both $t\bar{b}$ and $\bar{t}b$ production, and the t-channel process as "$tqb$," standing for both $tq\bar{b}$ and $\bar{t}\bar{q}b$. The next-to-leading order SM cross sections [2] are 0.9 ± 0.1 pb (s-channel) and 2.0 ± 0.3 pb (t-channel) for a top quark mass of 175 GeV. Single top quark production is very distinct from $t\bar{t}$ pair production since it comes from an electroweak $Wtb$ vertex instead of a strong $gtt$ one. Identifying single top quark events allows us to probe the $Wtb$ vertex in ways not accessible anywhere else. Figure 1 shows the main Feynman diagrams for single top quark production and decay at the Tevatron.

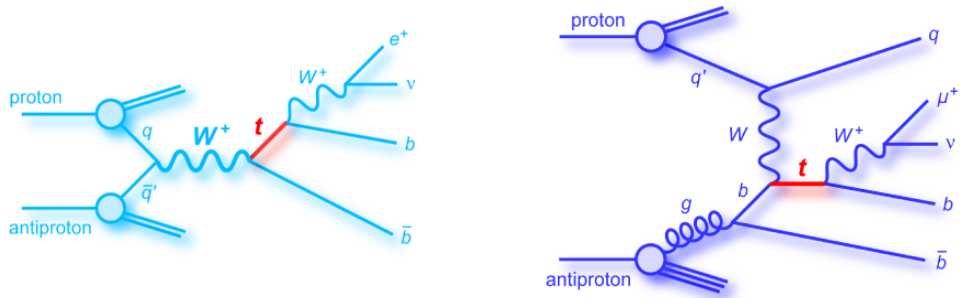

Figure 1: Principal tree-level Feynman diagrams for s-channel (left) and t-channel (right) single top quark production.

## 2. SELECTING SINGLE TOP QUARK EVENTS

The results presented here from the DØ collaboration use 0.9 fb$^{-1}$ of data, and 2.2 fb$^{-1}$ from the CDF collaboration, unless otherwise noted. This data was collected from 2002 to 2007. The signal signature is straightforward: a top quark is predicted to decay almost 100% of the time to a $W$ boson and a bottom quark; the experiments reconstruct the leptonic decays of the $W$ boson to an electron and neutrino (inferred from an imbalance in the total energy of the event, known as missing transverse energy, $\not{E}_T$) or a muon and neutrino. Thus, the final state consists of an isolated electron or muon, missing transverse energy, a $b$ jet from the top quark decay, and a second jet produced with the top quark. This is another $b$ jet for s-channel production, and a light-quark jet in the t-channel case. Occasionally in the t-channel, a second $b$ jet is produced from the splitting of the initial-state gluon with enough transverse momentum $p_T$ to be reconstructed. CDF required the electron or muon to have $p_T > 20$ GeV and pseudorapidity $|\eta| < 1.6$. DØ required $p_T > 15$ GeV and $|\eta| < 1.1$ for electrons, and $p_T > 18$ GeV and $|\eta| < 2.0$ for muons. CDF set the threshold for $\not{E}_T > 25$ GeV, whereas DØ set it at $> 15$ GeV. Both collaborations had so-called triangle cuts that reject events with low $\not{E}_T$ above these thresholds when very small or very large opening angles between the lepton or jets imply possible misreconstruction and thus mismeasured $\not{E}_T$. CDF required two or three jets with $p_T > 20$ GeV and $|\eta| < 2.8$. DØ required two, three, or four jets with $p_T(\text{jets } 1,2,3,4) > 25, 20, 15, 15$ GeV, and $|\eta(\text{jets } 1,2,3,4)| < 2.5, 3.4, 3.4, 3.4$. DØ included events with a fourth jet because there could have been an initial-state- or final-state-radiated quark or gluon. The background processes that these cuts selected were mainly from $W$+jets events at low jet multiplicity and $t\bar{t}$ pairs at high jet multiplicity, with small contributions from $Z$+jets, dibosons, and multijet events where one of the jets was misidentified as an electron, or a $b$ jet decayed to produce a muon that was misleadingly reconstructed as isolated from the jet. All the backgrounds (except multijets) and the signal events were simulated using Monte Carlo (MC) models: MadEvent (CDF) and CompHEP (DØ) for signals and Alpgen+Pythia for the backgrounds, followed by full detector simulations. The multijet backgrounds were modeled using data. Each experiment then applied sophisticated algorithms to identify ("tag") $b$ jets, and required one or two jets in the event to be tagged. Note that in this iteration of the analysis, DØ included the $Z$+jets and diboson backgrounds in the $W$+jets model. The event yields after selection are shown in Table I. Signal:background ratios ranged from 1:9 in the 2-jets/2-tags channels to 1:25 in the 3-jets/1-tag channels; before $b$ tagging, the ratio was 1:180. The acceptances for single top quark signals in all production and decay channels combined were 2.8% (CDF) and 3.2% (DØ) in the s-channel and 1.8% (CDF) and 2.1% (DØ) in the t-channel. Figure 2 shows distributions for the transverse mass of the reconstructed $W$ boson from CDF's analysis.

Table I: Event yields after selection. The color keys are used in subsequent plots in this paper.

| CDF preliminary 2.2 fb⁻¹ | | | DØ PRD 0.9 fb⁻¹ | | | |
|---|---|---|---|---|---|---|
| 1,2 b-tags | 2-jets | 3-jets | 1,2 b-tags | 2-jets | 3-jets | 4-jets |
| s-channel tb | 41 | 14 | s-channel tb | 16 | 8 | 2 |
| t-channel tqb | 62 | 18 | t-channel tqb | 20 | 12 | 4 |
| W+light-jets | 340 | 102 | W+light-jets | 119 | 43 | 12 |
| W+charm | 395 | 109 | W+charm | 151 | 85 | 23 |
| W+bottom | 462 | 141 | W+bottom | 261 | 120 | 35 |
| Z+jets | 27 | 11 | t$\bar{t}$→dileptons | 39 | 32 | 11 |
| Dibosons | 63 | 22 | t$\bar{t}$→lepton+jets | 20 | 103 | 143 |
| t$\bar{t}$ | 146 | 339 | Multijets | 95 | 77 | 29 |
| Multijets | 60 | 21 | | | | |
| Total prediction | 1,595 | 777 | Total prediction | 721 | 480 | 259 |
| Data | 1,535 | 712 | Data | 697 | 455 | 246 |

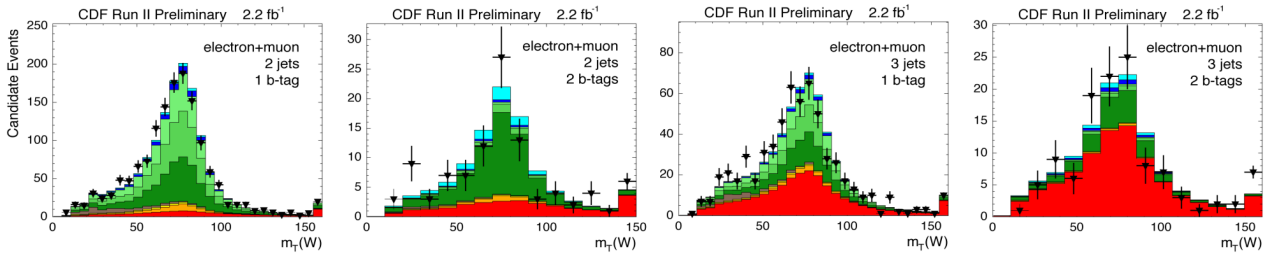

Figure 2: CDF's $M_T(W)$ distributions in single-tagged and double-tagged channels with two or three jets. (See Table I for color key.)

## 3. CROSS SECTION MEASUREMENTS

### 3.1. DØ's tb+tqb Analyses and Results

DØ completed a search for single top quark production at the end of 2006 that concluded with the first evidence (i.e., > 3 standard deviation significance) for *tb+tqb* production [3]. Three novel discriminant methods were applied to separate signal from background, with the data divided into 12 independent channels (electron or muon decay; 2, 3, or 4 jets; 1 or 2 *b*-tagged jets) to maximize the sensitivity. Boosted decision trees [4] were used to combine 49 variables that each had some separation between one or both signals and one or more background components. The output variable ranged from 0 (background-like) to 1 (signal-like) and was used in a Bayesian likelihood calculation to measure the signal cross section. Ensembles of pseudo-data selected from MC and data background events, with full systematic uncertainties included, were used to determine the significance of the measurement. Two other methods were also used to separate signal from background, Bayesian neural networks [5] and matrix elements [6], and a combination of all three methods with an improved signal significance of 3.6 standard deviations was published this year [7]. The measured *tb+tqb* cross section from the best linear unbiased estimate BLUE [8] method combination is 4.7 ± 1.3 pb. Figure 3 shows close-ups of the high ends of the three discriminants for all analysis channels combined. Table II shows the results for these three analyses and their combination.

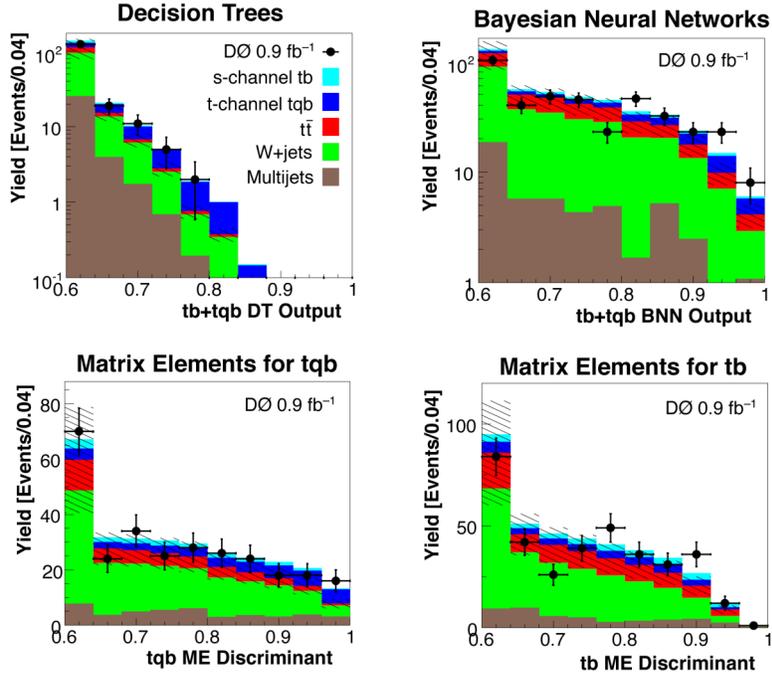

Figure 3: DØ's discriminant outputs for boosted decision trees (upper left), Bayesian neural networks (upper right) and matrix elements (lower row).

Table II: DØ's results for the *tb+tqb* cross section for a top quark mass of 175 GeV.

| *tb+tqb* Results | DØ PRD 0.9 fb$^{-1}$ | | | |
|---|---|---|---|---|
| | BDT | BNN | ME | BLUE Combination |
| **Expected significance** | 2.1 σ | 2.2 σ | 1.9 σ | 2.3 σ |
| **Observed significance** | 3.4 σ | 3.1 σ | 3.2 σ | 3.6 σ |
| **Cross section** | $4.9^{+1.4}_{-1.4}$ pb | $4.4^{+1.6}_{-1.4}$ pb | $4.8^{+1.6}_{-1.4}$ pb | $4.7^{+1.3}_{-1.3}$ pb |

## 3.2. CDF's tb+tqb Analyses and Results

The CDF collaboration has released several new measurements of the *tb+tqb* cross section over the past year. I present here results from the time of the HCP2008 conference that used 2.2 fb$^{-1}$ of data; ones that use 2.7 fb$^{-1}$ were released more recently, but are similar in nature and have not yet been combined for best precision. CDF used four different discrimination methods to separate signal from background: likelihood functions [9], neural networks [10], matrix elements [11], and boosted decision trees [12]. The latter three methods all have very similar high expected significance. CDF has combined [13] the first three results using a modified version of the BLUE method (allowing for iteration with asymmetric uncertainties) and also using a novel neural network method called "neuro-evolution of augmenting topologies (NEAT) [14] to obtain improved overall precision and significance. Figure 4 shows the output distributions from these discriminants and Table III shows the results separately and in combination.

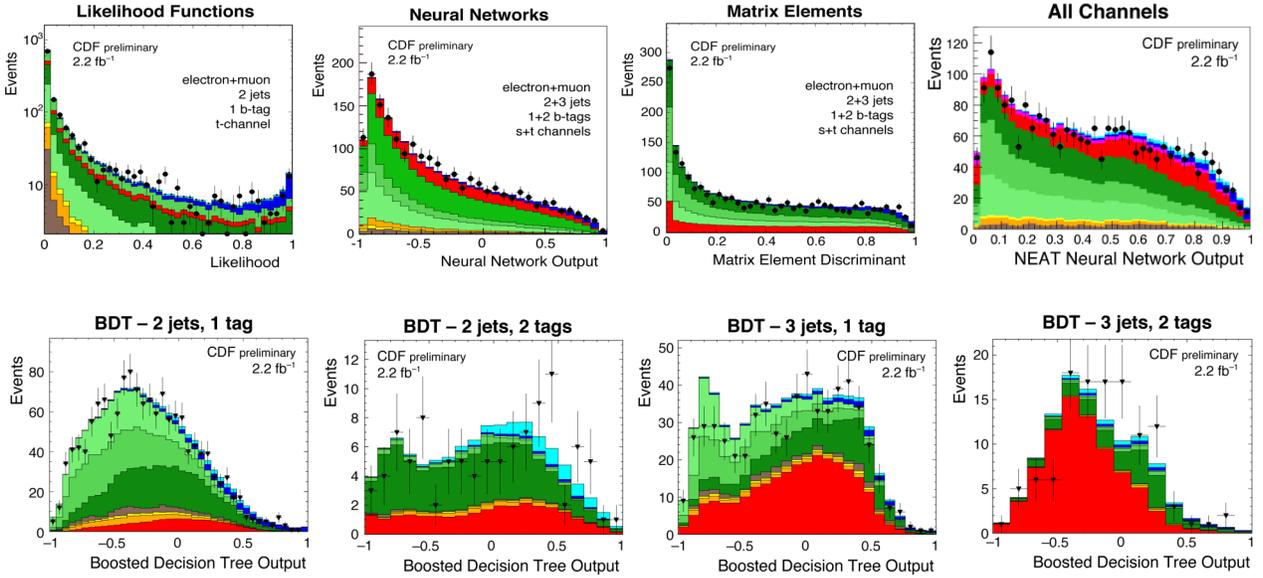

Figure 4: CDF's discriminant outputs for (upper row, left-to-right) likelihood functions, neural networks, matrix elements, and the NEAT neural network combination. The lower row shows the boosted decision trees outputs. (See Fig. 2 for color key.)

Table III: CDF's results for the *tb+tqb* cross section, for a top quark mass of 175 GeV.

| *tb+tqb* Results CDF preliminary 2.2 fb$^{-1}$ | LF | NN | ME | BLUE Combination | NEAT Combination | BDT |
|---|---|---|---|---|---|---|
| Expected significance | 3.4 σ | 4.4 σ | 4.5 σ | 4.7 σ | 5.1 σ | 4.6 σ |
| Observed significance | 2.0 σ | 3.2 σ | 3.4 σ | 3.7 σ | 3.7 σ | 2.8 σ |
| Cross section | $1.8^{+0.9}_{-0.8}$ pb | $2.0^{+0.9}_{-0.8}$ pb | $2.2^{+0.8}_{-0.7}$ pb | $2.1^{+0.7}_{-0.6}$ pb | $2.2^{+0.7}_{-0.7}$ pb | $1.9^{+0.8}_{-0.7}$ pb |

## 3.3. Separate Measurements of tb and tqb

Both the DØ and CDF collaborations have measured the cross sections for the s-channel *tb* and t-channel *tqb* processes separately. Two sets of measurements have been made. The first assume the SM cross section for the single top quark process not being measured. The second measurements do not include this assumption, and fit the cross sections to data and background model (DØ) or to signal and background templates (CDF) simultaneously.

DØ used boosted decision trees trained on each of these signals versus the background for the first measurement, and obtained $\sigma(p\bar{p} \to tb + X) = 1.0 \pm 0.9$ pb and $\sigma(p\bar{p} \to tqb + X) = 4.2^{+1.8}_{-1.4}$ pb [3]. These results are illustrated in Fig. 5. CDF used a likelihood function to separate *tb* from background, assuming the SM cross section for *tqb*. This analysis used 1.9 fb$^{-1}$ of data with exactly two jets, both *b* tagged. A small excess of data over background was seen and they set an upper limit at the 95% CL of $\sigma(p\bar{p} \to tb + X) < 2.8$ pb. The likelihood discriminant distribution is shown in Fig. 5.

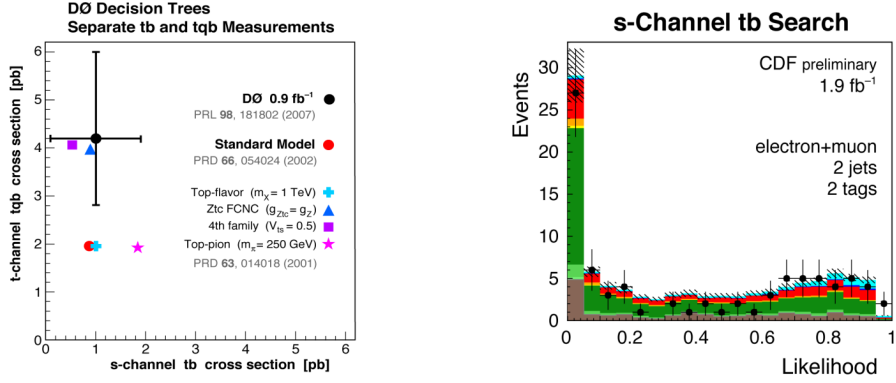

Figure 5: DØ's separate measurements of the two single top quark processes (left) and CDF's likelihood function used to set an upper limit on *tb* production (right).

DØ used the decision trees trained on *tb+tqb* signal and performed a 2-d fit, allowing the *tb* and *tqb* cross sections to both float. The results are $\sigma(p\bar{p} \to tb + X) = 0.9$ pb and $\sigma(p\bar{p} \to tqb + X) = 3.8$ pb [7], which are shown in Fig. 6. CDF used neural networks to perform a 2-d fit to templates for all signal and background processes, and measured $\sigma(p\bar{p} \to tb + X) = 1.6^{+0.9}_{-0.8}$ pb and $\sigma(p\bar{p} \to tqb + X) = 0.8^{+0.7}_{-0.8}$ pb, also shown in Fig. 6.

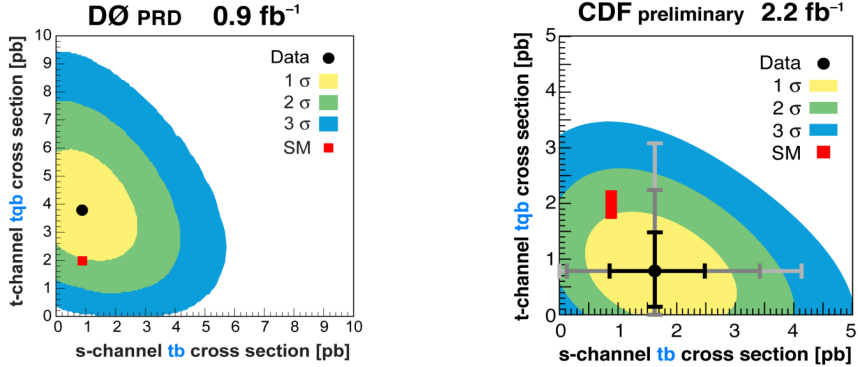

Figure 6: DØ's combined fit to the s- and t-channel cross sections (left) and CDF's similar measurement (right).

## 4. SINGLE TOP QUARK STUDIES

### 4.1. CKM Matrix Element |V$_{tb}$|

DØ used the *tb+tqb* cross section from the decision trees measurement to make the first direct measurement of the CKM matrix element $|V_{tb}|$ without assuming three generations (i.e., CKM unitarity) [3,7]. They found $|V_{tb}f_1^L| = 1.3 \pm 0.2$ where $f_1^L$ is an arbitrary left-handed form factor in the *Wtb* vertex (see Fig. 7), and $f_1^R$, $f_2^L$, $f_2^R$ are assumed to be zero. When $|V_{tb}|$ was restricted to lie between 0 and 1, then they found $0.68 \leq |V_{tb}| \leq 1$. The posterior density for $|V_{tb}|^2$ for the latter case is shown in Fig. 7. CDF has also measured $|V_{tb}|$, using the combination of their first three measurements of *tb+tqb*, restricting $|V_{tb}|$ to be $\leq 1$; they obtained $0.66 \leq |V_{tb}| \leq 1$. This result is also shown in Fig. 7.

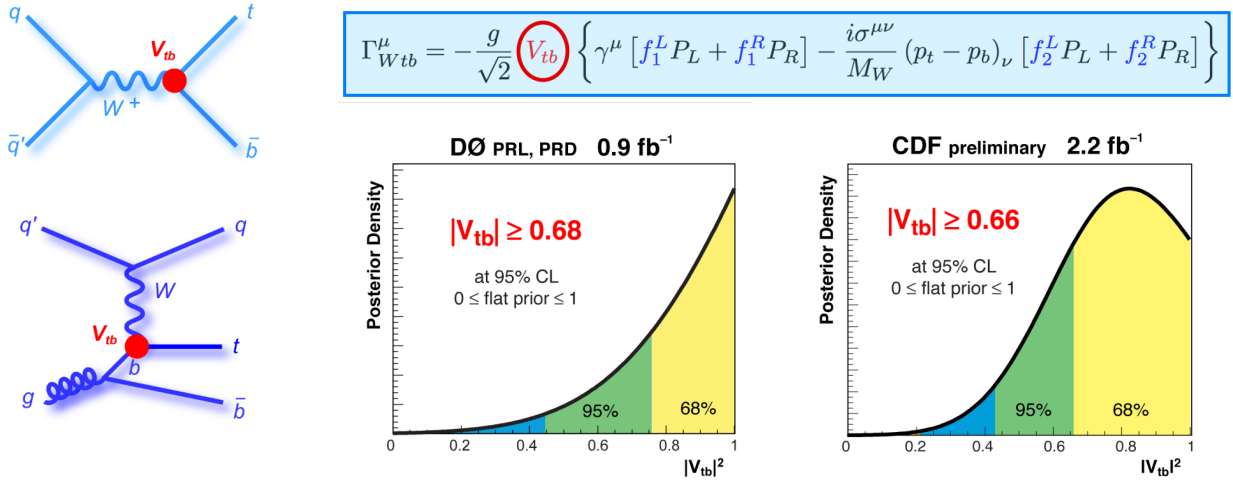

Figure 7: Feynman diagrams for single top quark production showing the *Wtb* vertex used to measure $|V_{tb}|$ (left), the *Wtb* coupling (upper right), and the posterior density distribution for $|V_{tb}|^2$ (DØ left, CDF right), used to obtain the limits on $|V_{tb}|$.

### 4.2. Flavor-Changing Neutral Currents

DØ used a 230 pb$^{-1}$ dataset to set limits on FCNC production of a single top quark with a light quark, charm quark, or gluon [15], by requiring exactly one *b*-tagged jet and combining 10 kinematic variables with a neural network for signal-background discrimination. The single top cross sections scale with $(\kappa_{gtu}/\Lambda)^4$ and $(\kappa_{gtc}/\Lambda)^4$. This measurement set limits on $\kappa_{gtu}$ and $\kappa_{gtc}$ that are 11x and 3x better than results from HERA: $\kappa_{gtu}/\Lambda < 0.037$ TeV$^{-1}$ and $\kappa_{gtc}/\Lambda < 0.15$ TeV$^{-1}$. The results are shown in Fig. 8. Since the HCP2008 conference, CDF has also completed a FCNC search using 2.2 fb$^{-1}$ with ~30% better limits [16] than DØ's published ones.

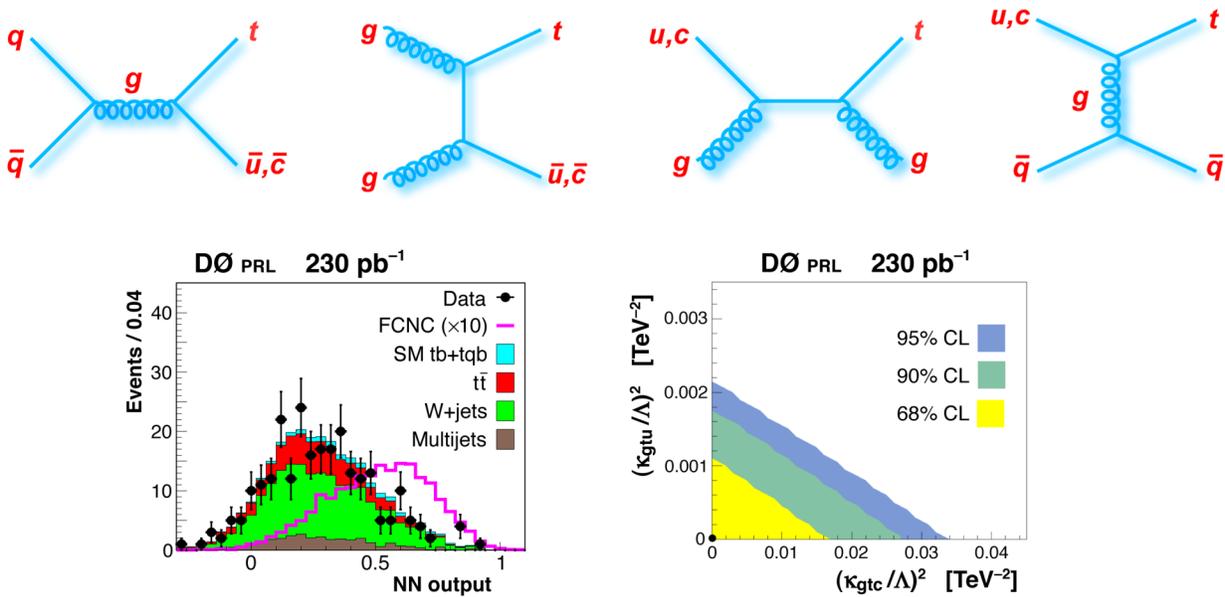

Figure 8: Flavor-changing neutral-current production of single top quarks: tree-level Feynman diagrams (upper row), neural network output distribution (left), and limit contours (right).

### 4.3. Heavy W′ Boson Resonances

The DØ collaboration has performed two searches for a heavy $W'$ boson s-channel resonance, using the 230 pb$^{-1}$ dataset [17] and an improved analysis with the 0.9 fb$^{-1}$ dataset [18] presented here. $W'$ bosons with either left-handed or right-handed couplings were considered, where the interference between the left-handed one and the SM s-channel $tb$ process was included in the signal model. DØ set limits for the case where the $W'$ could decay leptonically as well as hadronically (when $\nu_R$ was less massive than the $W'$) and when the $W'$ could only decay hadronically (when $\nu_R$ was more massive than the $W'$). DØ found $M(W'_L) > 731$ GeV, $M(W'_R) > 739$ GeV (low-mass $\nu_R$), and $M(W'_R) > 768$ GeV (high-mass $\nu_R$), at 95% CL. These results are illustrated in Fig. 9. The CDF collaboration performed a search for $W'$ production in Run I with 106 pb$^{-1}$ of data [19] and have a new search using 1.9 fb$^{-1}$ of data [20]. They assumed that for high-mass $W'$, the interference between left-handed bosons and the SM ones was negligible, and it was not included in their signal model. CDF found $M(W'_L) > 800$ GeV, $M(W'_R) > 800$ GeV (low-mass $\nu_R$), and $M(W'_R) > 825$ GeV (high-mass $\nu_R$), at 95% CL, shown in Fig. 9.

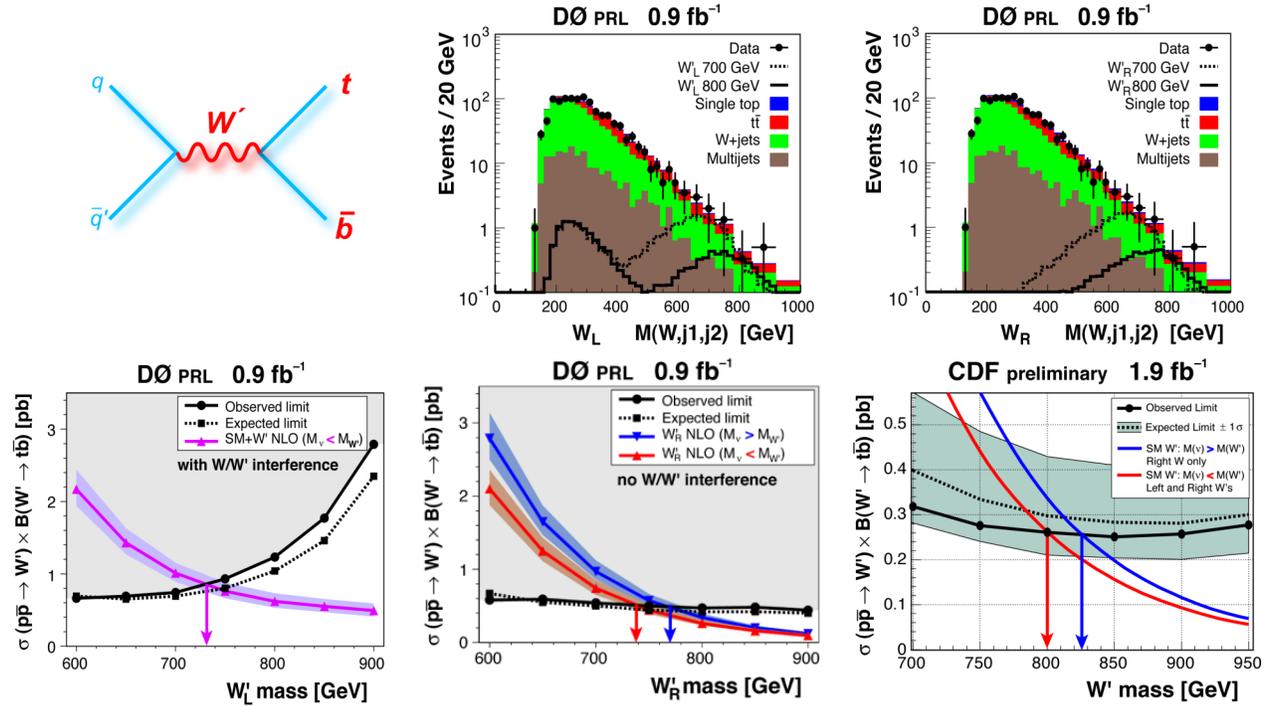

Figure 9: $W'$ boson resonance search: main Feynman diagram (upper left), DØ's invariant mass distributions showing left-handed (upper center) and right-handed (upper right) $W'$ signals, and limit plots (lower row) for DØ (left and center) and CDF (right).

### 4.4. Charged Higgs Boson Resonance

DØ has completed a new search using the 0.9 fb$^{-1}$ dataset for charged Higgs bosons that decay to $tb$ [21]. This was more difficult than the $W'$ search since the predicted cross sections in several two-Higgs-doublet models are much lower, so the mass range where there is sensitivity is also much lower, coinciding with the peak of the $W$+jets background. This is the first such search. They selected events with exactly two jets, one or two of them $b$ tagged, and performed a binned likelihood calculation using the invariant-mass distribution to obtain upper limits on the charged Higgs boson cross sections. Figure 10 shows the results from this analysis.

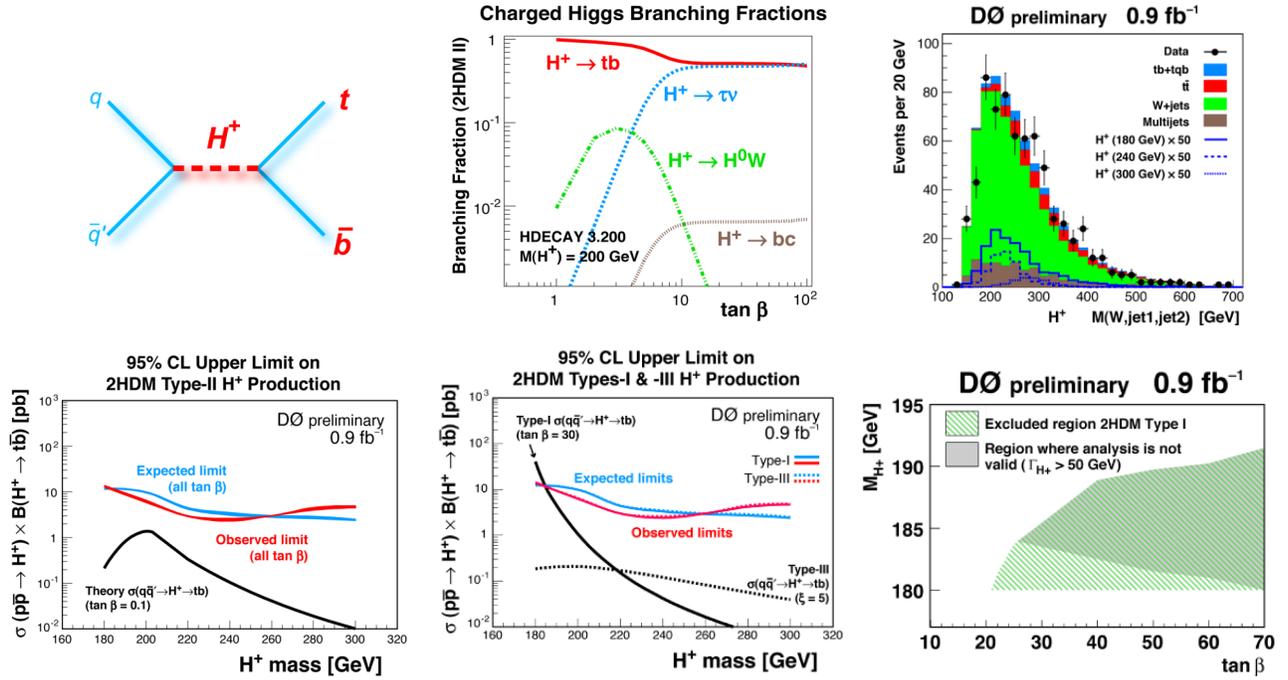

Figure 10: Charged Higgs boson resonance search: main Feynman diagram (upper left), branching fractions (upper center), invariant mass distribution (upper right), and limit plots (lower row).

### 4.5. Anomalous Wtb Couplings

DØ has performed an analysis using the 0.9 fb$^{-1}$ dataset to see whether the *Wtb* coupling is pure left-handed vector in form, or whether there are right-handed vector, or left- or right-handed tensor components [22]. They used boosted decision trees to discriminate between each type of signal and the background, and set upper limits on $|V_{tb}f|^2$ on pairs of the form factors $f_1^L, f_1^R, f_2^L, f_2^R$ (see Fig. 7). This measurement places the first limits on the *Wtb* tensor couplings $f_2^L$ and $f_2^R$. For the $f_1^L, f_2^L$ left-handed-couplings-only scenario, $|f_1^L|^2 = 1.4^{+0.6}_{-0.5}$ and $|f_2^L|^2 < 0.5$ at 95% CL. For the $f_1^L, f_1^R$ vector-couplings-only scenario, $|f_1^L|^2 = 1.8^{+1.0}_{-1.3}$ and $|f_1^R|^2 < 2.5$ at 95% CL. For the $f_1^L, f_2^R$ mixed left-right, vector-tensor scenario, $|f_1^L|^2 = 1.4^{+0.9}_{-0.8}$ and $|f_2^R|^2 < 0.3$ at 95% CL. The 2-d limit contours are shown in Fig. 11.

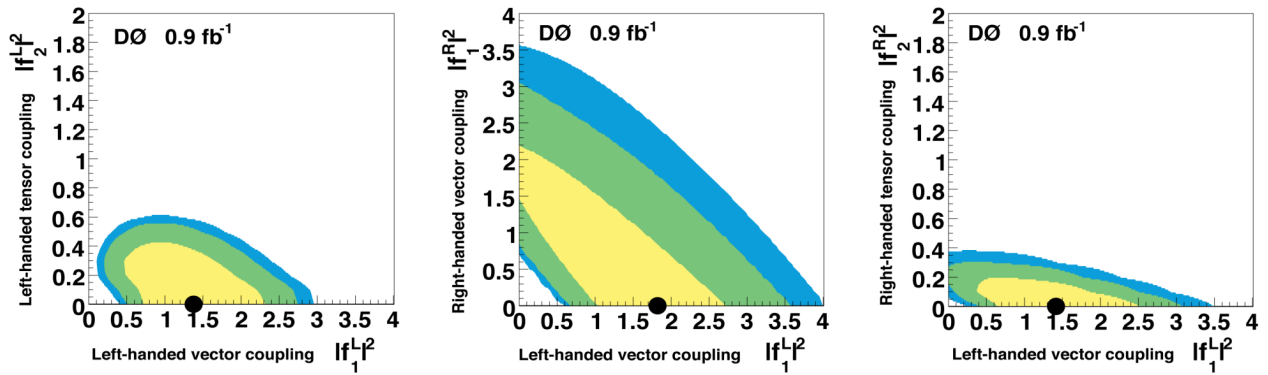

Figure 11: Measured coupling strengths (black solid circles), with 1, 2, 3 standard deviation limit regions for each anomalous coupling form factor squared versus the SM left-handed vector-coupling form factor squared.

## 5. OUTLOOK

A strong program of single top quark physics has been established over the last two years since evidence of its existence was found at the end of 2006. The measurements are challenging because the signal cross sections are rather small and the multi-component background is very large. DØ has analyzed 0.9 fb$^{-1}$ of data and measured a cross section for $tb+tqb$ of $4.7 \pm 1.3$ pb with $3.6\sigma$ significance [3,7]. CDF has analyzed 2.2 fb$^{-1}$ of data and measured the cross section at $2.2 \pm 0.7$ pb with $3.7\sigma$ significance [13]. Measurements have been made of the $tb$ and $tqb$ cross sections separately, of the CKM matrix element $|V_{tb}|$, and limits set on $tb$ production from $W'$ and $H^+$ resonances, and on anomalous $Wtb$ couplings. Figure 12 shows the cross section measurements from each analysis method, and the projected sensitivity of each experiment as more data is analyzed in the future.

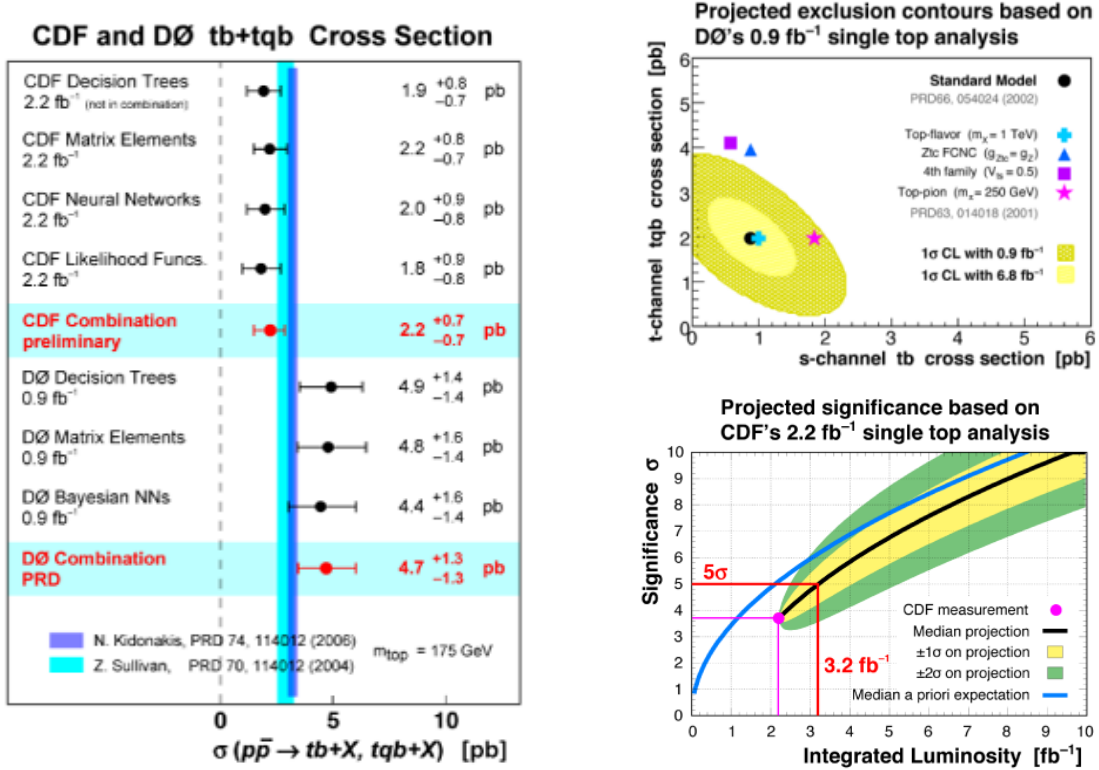

Figure 12: Summary of the cross section measurements for combined $tb+tqb$ production at the Tevatron (left), and projections for future significance based on the analyses presented here carried out with larger datasets (right).

### Acknowledgments

The author wishes to thank the symposium organizers for a very interesting meeting, and members of the single top working groups from the CDF and DØ collaborations for reviewing the talk material. Work supported by Department of Energy grant number DE-FG02-04ER40837.